# Adaptive sensing using deterministic partial Hadamard matrices

S. Haghighatshoar, E. Abbe and E. Telatar
EPFL, Switzerland
Email: {saeid.haghighatshoar, emmanuel.abbe, emre.telatar} @epfl.ch

*Abstract*—**This paper investigates the construction of deterministic matrices preserving the entropy of random vectors with a given probability distribution. In particular, it is shown that for random vectors having i.i.d. discrete components, this is achieved by selecting a subset of rows of a Hadamard matrix such that (i) the selection is deterministic (ii) the fraction of selected rows is vanishing. In contrast, it is shown that for random vectors with i.i.d. continuous components, no partial Hadamard matrix of reduced dimension allows to preserve the entropy. These results are in agreement with the results of Wu-Verdu on almost lossless analog compression. This paper is however motivated by the complexity attribute of Hadamard matrices, which allows the use of efficient and stable reconstruction algorithms. The proof technique is based on a polar code martingale argument and on a new entropy power inequality for integer-valued random variables.**

## I. Introduction

Information theory has extensively studied the lossless and lossy compression of discrete time signals into *digital* sequences. These problems are motivated by the model of Shannon, where an analog signal is first acquired, by sampling it at a high enough rate to preserve all of its information (Nyquist-Shannon sampling theorem), and *then* compressed. More recently, it was realized that proceeding to "joint sensing-compression" schemes can be beneficial. In particular, *compressed sensing* introduces the perspective that sparse signals can be compressively sensed to decrease measurement rate. As for joint source-channel coding schemes, one may wonder why this would be useful? Eventually, the signal is represented with the same amount of bits, so why would it be preferable to proceed jointly or separately? In a nutshell, if measurements are expensive (such as for example in certain bio-medial applications), then compressed sensing is beneficial.

From an information-theoretic perspective, compressed sensing can be viewed as a form of analog to analog compression. Namely, transforming a discrete time signal into a lower-dimensional discrete time signal over the reals, without "losing information". The key point being that, since measurements are analog, one may as well pack as much information in each measurement (whereas in the compression of discrete signals, a measurement on a larger alphabet is more expensive than a measurement in bits). However, compressing a vector in $\mathbb{R}^n$ into a vector in $\mathbb{R}^m$, $m < n$, without regularity constraints is not an interesting problem, since $\mathbb{R}^n$ and $\mathbb{R}^m$ have the same cardinality[1]. Hence, analog compression without any regulating conditions is trivial (as opposed to compression over finite fields).

Recently, [1] introduced a more reasonable framework to study analog compression from an information-theoretic perspective. By requiring the encoder to be linear and the decoder to be Lipschitz continuous, the fundamental compression limit is shown to be the Rényi information dimension. The setting of [1] also raises a new interesting problem: how to reach this limit with low-complexity schemes? In the same way that coding theory aims at approaching the Shannon limit with low-complexity schemes, it is a challenging problem to devise efficient schemes to reach the Rényi dimension. Indeed, in this analog framework, realizing measurements in a low complexity manner is at the heart of the problem: it is rather natural that the Rényi dimension is the fundamental limit irrespective of complexity considerations, but without a low-complexity scheme, one may not have any gain in proceeding with a joint compression-sensing approach. For example in the compressed sensing, with $O(k\log(n/k))$ instead of $O(k)$ measurements, $k$-sparse signals can be reconstructed using $l_1$ minimization, which is a convex optimization problem, rather than $l_0$ minimization, which is intractable [5], [6]. Hence, in general, complexity requirements may raise the measurement rate.

The scope of this paper is precisely to investigate what measurement rates can be achieved by taking into account the complexity of the sensing matrix, which in turn, influences the complexity of the reconstruction algorithm. Our goal is to consider signals that are memoryless and drawn from a distribution on $\mathbb{R}$, which may be purely atomic, purely continuous or mixed. This paper focuses on the purely atomic and purely continuous cases. It is legitimate to attempt reaching this goal by borrowing tools from coding theory, in particular from codes achieving least compression rates in the discrete setting. Our approach is based on using Hadamard matrices for the encoding and developing a counter-part of the polar technique [2], [3] with arithmetic over $\mathbb{R}$ (or $\mathbb{Z}$ for atomic distributions) rather than $\mathbb{F}_2$ or $\mathbb{F}_q$. The proof technique uses the martingale approach of polar codes and a new form of entropy power inequality for discrete distributions. Rigorous results are obtained and sensing matrix construction is deterministic. A nested property is also investigated which allows one to adapt the measurement rate to the sparsity level of the signal.

[1]Of course, such an approach is not practical in many regards: the approach would dramatically fail in the presence of noise, it would be problematic to implement nonlinear measurement, and it would be highly complex.

Recently, spatially-coupled LDPC codes have recently allowed to achieve rigorous results in coding theory. This approach has been exploited by [4], which proposes the use of spatially coupled matrices for sensing. In [4], the mixture case is covered and further analysis on the reconstruction algorithm is provided. However, the sensing matrix is still random. It is known that Hadamard matrices truncated randomly afford desirable properties for compressed sensing. In this paper, we show that by knowing the signal distribution, Hadamard matrices can be truncated deterministically and yet reach lower measurement rates.

## II. Related Work

Let $X_1, X_2, \ldots, X_N$, be i.i.d discrete random variables taking values in $\mathcal{X} = \{0, 1, \ldots, q-1\}$ with probability distribution $p_X$, where $q \in \mathbb{Z}_+$ and $N = 2^n$ for some positive integer $n$. We use the notation $a_i^j$ for the column vector $(a_i, a_{i+1}, \ldots, a_j)^t$ and set $a_i^j$ to null if $j < i$. We also define $[r] = \{i \in Z : 1 \le i \le r\}$.

In Arikan's source coding [3], $q$ is a prime number and the arithmetic is over $\mathbb{F}_q$. Defining $G = \begin{pmatrix} 1 & 1 \\ 0 & 1 \end{pmatrix}^{\otimes n}$, where $\otimes$ denotes the Kronecker product, $Y_1^N = G X_1^N$ over $\mathbb{F}_q$, and $H_i = H(Y_i | Y_1^{i-1})$, $i \in [N]$, as the conditional entropy of $Y_i$ given $Y^{i-1}$, one obtains

$$\sum_{i=1}^N H_i = H(Y_1^N) = H(X_1^N) = N H(X).$$

The polarization phenomenon states that for any $\delta > 0$ and as $n$ goes to infinity

$$\frac{\#\{i \in [N] : H_i \in (\delta, 1-\delta)\}}{N} \to 0,$$

where $H(X)$ denotes the entropy of $X$ in base $q$. This implies that for large $n$, the values $H_i$, $i \in [N]$, have polarized to 0 or 1. This provides a compression scheme achieving the least compression rate, since for every $\delta \in (0, 1)$

$$\frac{\#\{i \in [N] : H_i \in (1-\delta, 1]\}}{N} \to H(X). \tag{1}$$

From another point of view, every $Y_i$ is associated with the $i$-th row of the matrix $G$ and (1) indicates that the "measurement" rate required to extract the informative components is close to the entropy of the source $H(X)$ for large $N$. This gives a good "measurement matrix" for a given distribution over $\mathbb{F}_q$.

In signal acquisition, measurements are analog. Hence, one can consider $Y_1^N = G X_1^N$ with arithmetic over the real field and investigate if any "polarization phenomenon" occurs. The difference is that, in this case, the measurement alphabet is unbounded. In particular, the $H_i$ values are not bounded above. We will show in Theorem 1 that instead of a polarization phenomenon, where two extremal states survive, an *absorption phenomenon* occurs where

$$\frac{\#\{i \in [N] : H_i > \delta\}}{N} \to 0,$$

as $N$ becomes larger, i.e., the measurement rate tends to 0.

## III. Problem Statement

**Definition 1** (Restricted iso-entropy property)**.** Let $X_1^N$ be discrete i.i.d. random variables with a marginal distribution $p_X$ supported on a finite set. The family $\{\Phi_N\}$ of measurement matrices, where $\Phi_N$ has dimension $m_N \times N$, is $\epsilon$-REP$(p_X)$ with measurement rate $\rho$ if

$$\frac{H(X_1^N | \Phi_N X_1^N)}{N} < \epsilon \tag{2}$$

and

$$\limsup_{N \to \infty} \frac{m_N}{N} = \rho.$$

In general, the labeling $N$ can be any subsequence of $\mathbb{Z}_+$. We will consider $N = 2^n, n \in \mathbb{Z}_+$. Also note that (2) is equivalent to

$$H(\Phi X_1^N) \le H(X_1^N) < H(\Phi X_1^N) + N\epsilon, \text{ when } X_1^N \overset{\text{iid}}{\sim} p_X,$$

which is similar in form to the RIP condition [5], [6], replacing energy ($l_2$ norm) with entropy, and sparsity with a probabilistic characterization (which may or may not relate to sparsity).

**Definition 2.** Let $X_1^N$ be continuous (or mixture) random variables with probability distribution $p_X$. The family of measurement matrices $\{\Phi_N\}$ of dimension $m_N \times N$ is $(\epsilon, \gamma)$-REP$(p_X)$ with measurement rate $\rho$ if

1) there exists a single letter quantizer $Q : \mathbb{R} \to \mathbb{Z}$ such that M.M.S.E. of $X$ given $Q(X)$ is less that $\gamma$,
2) for any $N$,
$$\frac{H(Q(X_1^N) | \Phi_N X_1^N)}{N} < \epsilon,$$
where $Q(X_1^N) = (Q(X_1), Q(X_2), \ldots, Q(X_N))^t$,
3) $$\limsup_{N \to \infty} \frac{m_N}{N} = \rho.$$

We address the following questions in this paper:

1) Given a probability distribution $p_X$ over a finite set, and $\epsilon > 0$, is there a family of measurement matrices that is $\epsilon$-REP and has measurement rate $\rho$? What is the set of all possible $(\epsilon, \rho)$ pairs? Is it possible to construct a near optimal family of truncated Hadamard matrices with a minimal measurement rate? How is the truncation adapted to the distribution $p_X$?
2) Is it possible to obtain an asymptotic measurement rate below 1 for continuous distributions?

*Remark* 1. The RIP notion introduced in [5], [6] is useful in compressed sensing, since it guarantees a stable $l_2$-recovery. We will consider truncated Hadamard matrices satisfying REP condition and since they have a Kronecker structure, we will obtain a low-complexity reconstruction algorithm. This part is however not emphasized in this paper, and we mainly focus on the construction of the truncated Hadamard matrices. Section VI provides numerical simulations of a divide and conquer ML decoding algorithm and illustrates the robustness of the recovery to noise. In a future work, we will investigate the use of a recovery algorithm à la [4].

## IV. Main Results

The main results of this paper are summarized here.

**Definition 3.** Let $\{J_N = \begin{pmatrix} 1 & 1 \\ -1 & 1 \end{pmatrix}^{\otimes n}, N = 2^n, n \in \mathbb{Z}_+\}$ be the family of Hadamard matrices. Suppose $X_1^N$ are i.i.d. random variables with distribution $p_X$ over a finite subset of $\mathbb{Z}$. Let $Y_1^N = J_N X_1^N$ and define $H_i = H(Y_i|Y_1^{i-1})$ and $m_N = \#\{i \in [N] : H_i > \epsilon\}$. The $(\epsilon, p_X)$-truncated Hadamard family $\{\bar{J}_N\}$, is the set of matrices of dimension $m_N \times N$ obtained by selecting those rows of $J_N$ with $H_i > \epsilon$.

**Theorem 1** (Absorption phenomenon). *Let $X$ be a discrete random variable with a probability distribution $p_X$ supported on a finite subset of $\mathbb{Z}$. For a fixed $\epsilon > 0$, the family of $(\epsilon, p_X)$-truncated Hadamard matrices $\{\bar{J}_N, N = 2^n, n \in \mathbb{Z}_+\}$ (defined above) are $\epsilon$-REP$(p_X)$ with measurement rate 0. In other words,*

$$\limsup_{N\to\infty} \frac{m_N}{N} = 0.$$

*Remark* 2. Although all of the measurement matrices $\bar{J}_N$ are constructed by truncating the matrices $J_N$, the order and number of the selected rows, $m_N$, to construct $\bar{J}_N$ depends on the distribution $p_X$.

The proof idea is to construct a conditional entropy martingale process similar to [3] which is bounded from below and hence converges almost surely. Then, the following "entropy power inequality" result which we prove in Subsection V-A is used to show the convergence to 0.

**Theorem 2** (An EPI over $\mathbb{Z}$). *For every probability distribution $p$ over $\mathbb{Z}$,*

$$H(p \star p) - H(p) \geq g(H(p)). \tag{3}$$

*where $g : \mathbb{R}_+ \to \mathbb{R}_+$ is strictly increasing, $\lim_{x\to\infty} g(x) = \frac{1}{8\log(2)}$ and $g(x) = 0$ if and only if $x = 0$.*

*Remark* 3. This theorem complements the work in [7] to obtain an entropy power inequality for discrete random variables.

For continuous distributions, and for any fixed distortion $\gamma$, the measurement rate approaches 1 as $\epsilon$ tends to 0. This result has been shown in [1] in a more general context. We recover this result in our setting for the case of a uniform distribution over $[-1, 1]$ which is proved in the Appendix.

**Lemma 1.** *Let $p_U$ be the uniform distribution over $[-1, 1]$ and let $Q : [-1, 1] \to \{0, 1, \dots, q-1\}$ be a uniform quantizer for $X$ with M.M.S.E. less than $\gamma$. Assume that $\{\Phi_N\}$ is a family of full rank measurement matrices of dimension $m_N \times N$. If $\{\Phi_N\}$ is $(\epsilon, \gamma)$-REP$(p_U)$, then the measurement rate, $\rho$, goes to 1 as $\epsilon$ tends to 0.*

## V. Proof Overview

### A. An EPI for discrete random variables

In this section we prove the EPI result stated in Theorem 2. If $p$ is a probability distribution over a finite set $\{0, 1, 2, \dots, q-1\}$, then from the continuity and strict concavity of the entropy function, there is always a guaranteed gap between $H(p \star p)$ and $H(p)$ and the gap is 0 if and only if $H(p) = 0$. Theorem 2 shows that this gap is uniformly bounded away from 0.

If $X$ and $Y$ are two real valued, continuous and independent random variables, then

$$2^{2h(X+Y)} \geq 2^{2h(X)} + 2^{2h(Y)}, \tag{4}$$

where $h$ denotes the differential entropy. Equality holds if and only if $X$ and $Y$ are Gaussian random variables. If $X$ and $Y$ have the same density $p$, then (4) becomes

$$h(p \star p) \geq h(p) + \frac{1}{2},$$

which implies a guaranteed increase of the differential entropy for i.i.d. random variables. For this reason we call (3) an EPI for discrete random variables.

**Lemma 2.** *Let $c > 0$ and suppose $p$ is a probability measure over $\mathbb{Z}$ such that $H(p) = c$. Then, for any $i \in \mathbb{Z}$,*

$$H(p \star p) - c \geq cp_i - (1 + p_i)h_2(p_i),$$

*where $h_2(x) = -x\log_2(x) - (1-x)\log_2(1-x)$ is the binary entropy function and $p_i$ denotes the probability of $i$.*

*Proof:* For a finite positive measure $v$ on $\mathbb{Z}$, define $H(v) = -\sum_{i\in\mathbb{Z}} v_i \log v_i$. Note that for $\gamma > 0$, we have

$$H(\gamma v) = L(\gamma) + \gamma H(v),$$

where $L(\gamma) = -\gamma \log \gamma$. Let $i \in \mathbb{Z}$, $x = p_i$ and let us write

$$p = x\delta_i + (1-x)\nu,$$

where $\nu$ is a probability measure on $\mathbb{Z} \setminus \{i\}$. Note that

$$H(p) = h_2(x) + (1-x)H(\nu) = c,$$

where $h_2$ denotes the binary entropy function. We also have

$$p \star p = x^2\delta_{2i} + 2x(1-x)\nu_i + (1-x)^2\nu \star \nu,$$

where for $k \in \mathbb{Z}$, $\nu_i(k) = \nu(k-i)$. By concavity of the entropy,

$$\begin{aligned} H(p \star p) &\geq 2x(1-x)H(\nu) + (1-x)^2 H(\nu \star \nu) \\ &\geq (1-x^2)H(\nu) \\ &= (1+x)(c - h_2(x)). \end{aligned}$$

Hence, $H(p \star p) - c \geq cx - (1+x)h_2(x)$. ■

**Lemma 3.** *Let $c > 0$, $0 < \alpha < \frac{1}{2}$ and $n \in \mathbb{Z}$. Assume that $p$ is a probability measure on $\mathbb{Z}$ such that $\alpha \leq p((-\infty, n]) \leq 1-\alpha$ and $H(p) = c$, then*

$$\|p \star p_1 - p \star p_2\|_1 \geq 2\alpha,$$

*where* $p_1 = \frac{1}{p((-\infty,n])} p|_{(-\infty,n]}$ *and* $p_2 = \frac{1}{p([n+1,\infty))} p|_{[n+1,\infty)}$ *are scaled restrictions of* $p$ *to* $(-\infty,n]$ *and* $[n+1,\infty)$ *respectively.*

*Proof:* Let $\alpha_1 = p((-\infty,n])$ and $\alpha_2 = p([n+1,\infty)) = 1-\alpha_1$. Note that $p = \alpha_1 p_1 + \alpha_2 p_2$. We distinguish two cases $\alpha_1 \le \frac{1}{2}$ and $\alpha_1 > \frac{1}{2}$. If $\alpha_1 \le \frac{1}{2}$ then we have

$$\begin{aligned}
&\|p \star p_1 - p \star p_2\| \\
&= \|\alpha_1 p_1 \star p_1 - (1-\alpha_1) p_2 \star p_2 + (1-2\alpha_1) p_1 \star p_2\|_1 \\
&\ge \|\alpha_1 p_1 \star p_1 - (1-\alpha_1) p_2 \star p_2\|_1 - (1-2\alpha_1)\|p_1 \star p_2\|_1 \\
&= \alpha_1 + (1-\alpha_1) - (1-2\alpha_1) = 2\alpha_1 \ge 2\alpha,
\end{aligned}$$

whereas if $\alpha_1 > \frac{1}{2}$ we have

$$\begin{aligned}
&\|p \star p_1 - p \star p_2\| \\
&= \|\alpha_1 p_1 \star p_1 - (1-\alpha_1) p_2 \star p_2 + (1-2\alpha_1) p_1 \star p_2\|_1 \\
&\ge \|\alpha_1 p_1 \star p_1 - (1-\alpha_1) p_2 \star p_2\|_1 - (2\alpha_1 - 1)\|p_1 \star p_2\|_1 \\
&= \alpha_1 + (1-\alpha_1) - (2\alpha_1 - 1) = 2(1-\alpha_1) \ge 2\alpha,
\end{aligned}$$

where we used the triangle inequality, $1-\alpha_1 > \alpha$ and the fact that $p_1 \star p_1$ and $p_2 \star p_2$ have non-overlapping supports, so the $\ell_1$-norm of the sum is equal to sum of the corresponding $\ell_1$-norms. ■

**Lemma 4.** *Assuming the hypotheses of Lemma 3,*

$$H(p \star p) - c \ge \frac{\alpha^2}{2\log(2)} \|p \star p_1 - p \star p_2\|_1^2.$$

*Proof:* Let $\alpha_1$ and $\alpha_2$ be the same as in Lemma 3. Let $\nu_1 = p_1 \star p$, $\nu_2 = p_2 \star p$, and for $x \in [0,1]$, define $\mu_x = x\nu_1 + (1-x)\nu_2$ and $f(x) = H(\mu_x)$. One obtains

$$\begin{aligned}
f'(x) &= -\sum (\nu_{1i} - \nu_{2i}) \log_2(\mu_{xi}), \\
f''(x) &= -\frac{1}{\log(2)} \sum \frac{(\nu_{1i} - \nu_{2i})^2}{\mu_{xi}} \le 0.
\end{aligned}$$

Hence $f(x)$ is a concave function of $x$. Moreover,

$$\begin{aligned}
f'(0) &= \quad D(\nu_1 \| \nu_2) + H(\nu_1) - H(\nu_2), \\
f'(1) &= -D(\nu_2 \| \nu_1) + H(\nu_1) - H(\nu_2).
\end{aligned}$$

Since $p_1$ and $p_2$ have separate supports, there is $i, j$ such that $\nu_{1i} = 0, \nu_{2i} > 0$ and $\nu_{1j} > 0, \nu_{2j} = 0$. Hence $D(\nu_1 \| \nu_2)$ and $D(\nu_2 \| \nu_1)$ are both equal to infinity. In other words,

$$\begin{aligned}
f'(0) &= +\infty, \\
f'(1) &= -\infty.
\end{aligned}$$

Hence the unique maximum of the function must happen between 0 and 1. Assume that for fixed $\nu_1$ and $\nu_2$, $x^\star$ is the maximizer. If $0 < \alpha_1 \le x^\star$ then

$$\alpha_1 f'(\alpha_1) = \sum \alpha_1 (\nu_{2i} - \nu_{1i}) \log_2(\mu_{\alpha_1 i}) \ge 0,$$

which implies that

$$\begin{aligned}
f(\alpha_1) &= -\sum \mu_{\alpha_1 i} \log_2(\mu_{\alpha_1 i}) \\
&= -\sum (\nu_{2i} + \alpha_1(\nu_{1i} - \nu_{2i})) \log_2(\mu_{\alpha_1 i}) \\
&\ge -\sum \nu_{2i} \log_2(\mu_{\alpha_1 i}) \\
&= H(\nu_2) + D(\nu_2 \| \mu_{\alpha_1}) \\
&\ge H(p) + \frac{1}{2\log(2)} \|\nu_2 - \mu_{\alpha_1}\|_1^2 \\
&= H(p) + \frac{\alpha_1^2}{2\log(2)} \|\nu_1 - \nu_2\|_1^2,
\end{aligned}$$

where we used Pinsker's inequality

$$D(r \| s) \ge \frac{1}{2\log(2)} \|r - s\|_1^2.$$

Similarly, we can show that if $x^\star \le \alpha_1 \le 1$ then

$$f(\alpha_1) \ge H(p) + \frac{(1-\alpha_1)^2}{2\log(2)} \|\nu_1 - \nu_2\|_1^2.$$

As $\alpha \le \alpha_1 \le 1 - \alpha$ and $\alpha \le \frac{1}{2}$ it results that

$$\begin{aligned}
H(p \star p) &= H(\alpha_1 p \star p_1 + (1-\alpha_1) p \star p_2) \\
&= f(\alpha_1) \\
&\ge H(p) + \frac{\alpha^2}{2\log(2)} \|\nu_1 - \nu_2\|_1^2 \\
&\ge c + \frac{\alpha^2}{2\log(2)} \|\nu_1 - \nu_2\|_1^2.
\end{aligned}$$

■

**Lemma 5.** *Assuming the hypotheses of Lemma 3,*

$$H(p \star p) - c \ge \frac{2\alpha^4}{\log(2)}.$$

*Proof:* Combine Lemma 3 and 4. ■

*Proof of Theorem 2:* Suppose that $p$ is a distribution over $\mathbb{Z}$ with $H(p) = c$. Set $y = \|p\|_\infty$. It is easy to see that $y \ge 2^{-c}$. Also there is an $\alpha \ge \frac{1-y}{2}$ and an integer $n$ such that $\alpha \le p((-\infty,n]) \le 1-\alpha$. Using Lemma 2 and Lemma 5, it results that $H(p \star p) - c \ge t(c)$ where

$$t(c) = \min_{y \in [2^{-c},1]} \max\left(\frac{(1-y)^4}{8\log(2)}, cy - (1+y)h_2(y)\right).$$

For simplicity we consider

$$g(c) = \min_{y \in [0,1]} \max\left(\frac{(1-y)^4}{8\log(2)}, cy - (1+y)h_2(y)\right),$$

which is less than or equal to $t(c)$. It is easy to check that $g(c)$ is a continuous function of $c$. The monotonicity of $g$ follows from the fact that $cy - (1+y)h_2(y)$ is an increasing function of $c$ for every $y \in [0,1]$. For strict positivity, note that $(1-y)^4$ is strictly positive for $y \in [0,1)$ and it is 0 when $y = 1$, but $\lim_{y \to 1} cy - (1+y)h_2(y) = c$. Hence for $c > 0$, $g(c) > 0$. If $c = 0$ then

$$\max\left(\frac{(1-y)^4}{8\log(2)}, cy - (1+y)h_2(y)\right) = \frac{(1-y)^4}{8\log(2)}$$

and its minimum over $[0,1]$ is 0.

For asymptotic behavior, note that at $y = 0$, $cy - (1 + y)h_2(y) = 0$ and $\frac{(1-y)^4}{8\log(2)} = \frac{1}{8\log(2)}$. Hence, from continuity, it results that $g(c) \le \frac{1}{8\log(2)}$ for any $c \ge 0$. Also for any $\epsilon > 0$ there exists a $c_0$ such that for any $\epsilon < y \le 1$, $cy - (1 + y)h_2(y) \ge \frac{1}{8\log(2)}$. Thus for any $\epsilon > 0$ there is a $c_0$ such that for $c > c_0$, the outer minimum over $y$ in $g(c)$ is achieved on $[0, \epsilon]$. Hence, for any $c > c_0$, $g(c) \ge \frac{(1-\epsilon)^4}{8\log(2)}$. This implies that for every $\epsilon > 0$,

$$\frac{1}{8\log(2)} \ge \limsup_{c\to\infty} g(c) \ge \liminf_{c\to\infty} g(c) \ge \frac{(1-\epsilon)^4}{8\log(2)},$$

and $\lim_{c\to\infty} g(c) = \frac{1}{8\log(2)}$. ■

Figure 1 shows the EPI gap. As expected, for large values of $H(p)$, the gap approaches the asymptotic value $\frac{1}{8\log(2)}$. This is very similar to the EPI bound obtained for continuous random variables an we believe that one can improve this asymptotic bound to achieve $\frac{1}{2}$.

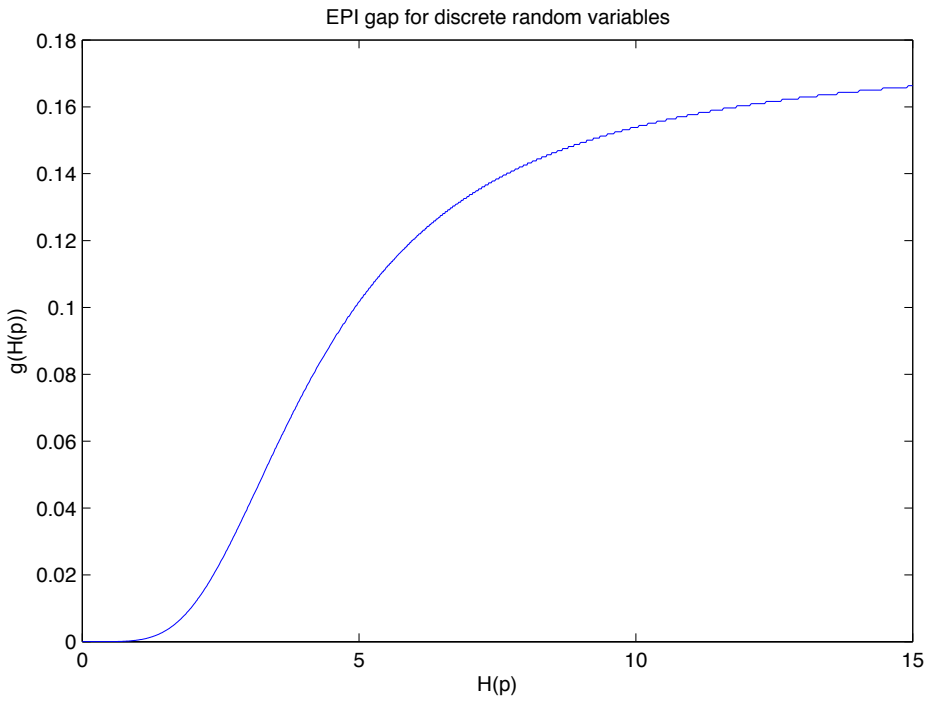

Fig. 1: EPI gap for discrete random variables

### B. Conditional Entropy Martingale

Assume that $X_1^N$, $N = 2^n$, $n \in \mathbb{Z}_+$, is a set of i.i.d. random variables with probability distribution $p_X$ over a finite subset of $\mathbb{Z}$. Let $Y_1^N = J_N X_1^N$, where $J_N = \begin{pmatrix} 1 & 1 \\ -1 & 1 \end{pmatrix}^{\otimes n}$ is the Hadamard matrix of dimension $N$ and let $H_i = H(Y_i|Y_1^{i-1}), i \in [N]$, be the conditional entropy values.

**Lemma 6.** *Let $X_1^N$ be as in the previous part and let $Z_1^N = G_N X_1^N$, where $G_N = \begin{pmatrix} 1 & 1 \\ 0 & 1 \end{pmatrix}^{\otimes n}$. Assume that $\tilde{H}_i = H(Z_i|Z_1^{i-1})$, then $H_i = \tilde{H}_i$, $i \in [N]$.*

*Remark* 4. The only point of Lemma 6 is that in application, it is preferred to use $J$ because the rows of $J$ are orthogonal to one another. For simplicity of the proof, we use $G$ matrices and relate to the polar code notations [2], [3].

*Proof:* We prove by induction over $n$ and consider the fact that $J_N$ and $G_N$ have similar recursive structure as a function of $J_{\frac{N}{2}}$ and $G_{\frac{N}{2}}$. For simplicity, we prove the lemma in a more general case. Assume that $Z_i, Y_i$, $i \in [N]$, are as introduced before. Suppose $\mathcal{O}$ is a random element and redefine $H_i = H(Y_i|Y_1^{i-1}, \mathcal{O})$ and $\tilde{H}_i = H(Z_i|Z_1^{i-1}, \mathcal{O})$. We prove that $H_i = \tilde{H}_i$. By putting $\mathcal{O}$ equal to null, we obtain the proof for the lemma. For $n = 1$ we have

$$\begin{aligned} H_1 &= H(Y_1|\mathcal{O}) \\ &= H(X_1 + X_2|\mathcal{O}) \\ &= H(Z_1|\mathcal{O}) \\ &= \tilde{H}_1. \end{aligned}$$

We also have

$$\begin{aligned} H(Y_1, Y_2|\mathcal{O}) &= H(X_1, X_2|\mathcal{O}) \\ &= H(Z_1, Z_2|\mathcal{O}). \end{aligned}$$

Hence, from the chain rule for conditional entropy we obtain that $H_2 = \tilde{H}_2$. Now assume that we have the result for all $n \le m$ and we prove it for $n = m + 1$. For simplicity, let us define the following notations

$$\begin{aligned} &V_1^{(m)} = X_1^{2^m} &&, \quad V_2^{(m)} = X_{2^m+1}^{2^{m+1}}, \\ &A_m = J_{2^m} &&, \quad B_m = G_{2^m}, \\ &R = B_m V_2^{(m)} &&, \quad S = A_m V_2^{(m)}, \\ &T = A_m(V_2^{(m)} - V_1^{(m)}). \end{aligned}$$

From the recursive structure of $J$ and $G$ matrices, the first $2^m$ components of $Z_1^{2^{m+1}}$ and $Y_1^{2^{m+1}}$ are equal to $A_m(V_1^{(m)} + V_2^{(m)})$ and $B_m(V_1^{(m)} + V_2^{(m)})$ respectively. The components of $V_1^{(m)} + V_2^{(m)}$ are i.i.d. random variables. Hence, using the induction hypothesis we obtain that the first $2^m$ components of $H_i$ and $\tilde{H}_i$ are equal. Now we prove that for $i = 2^m + 1, \ldots, 2^{m+1}$, they are also equal. For $2^m + 1 \le i \le 2^{m+1}$, setting $j = i - 2^m$ we have

$$\begin{aligned} H_i &= H(R_j|B_m(V_1^{(m)} + V_2^{(m)}), R_1^{j-1}, \mathcal{O}) \\ &= H(R_j|V_1^{(m)} + V_2^{(m)}, R_1^{j-1}, \mathcal{O}), \end{aligned}$$

and

$$\begin{aligned} \tilde{H}_i &= H(T_j|A_m(V_1^{(m)} + V_2^{(m)}), T_1^{j-1}, \mathcal{O}) \\ &= H(T_j|V_1^{(m)} + V_2^{(m)}, T_1^{j-1}, \mathcal{O}) \\ &= H(S_j|V_1^{(m)} + V_2^{(m)}, S_1^{j-1}, \mathcal{O}), \end{aligned}$$

where we used the invertibility of $A_m$ and $B_m$. Setting $\mathcal{O}' = \{V_1^{(m)} + V_2^{(m)}, \mathcal{O}\}$ and using the induction hypothesis, we obtain that for $2^m + 1 \le i \le 2^{m+1}$, $H_i = \tilde{H}_i$. Hence the induction proof is complete. Now setting $\mathcal{O}$ equal to null, we obtain the proof for Lemma 6. ■

Notice that we can represent $G_N$ in a recursive way. Let us define two binary operation $\oplus$ and $\ominus$ as follows

$$\begin{aligned} \ominus(a, b) &= a + b \\ \oplus(a, b) &= b, \end{aligned}$$

where $+$ is the usual integer addition. It is easy to see that we can do the multiplication by $G_N$ in a recursive way. Figure 2 shows a simple case for $G_4$. The $-$ or $+$ sign on an arrow shows that the result for that arrow is obtained by applying a $\ominus$ or $\oplus$ operation to two input operands.

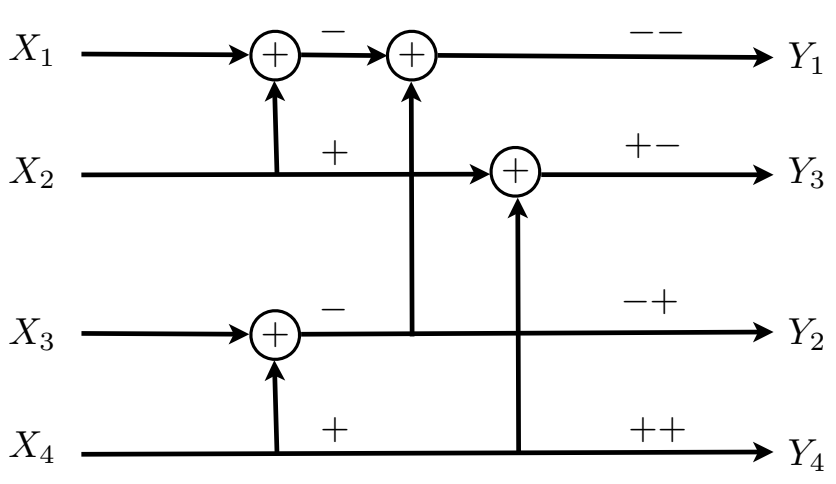

Fig. 2: Recursive structure for multiplication by $G_4$

If we consider a special output $Y_m$, there are a sequence of $\oplus$ and $\ominus$ operations on the input random variables which result in $Y_m$. An easy way to find this sequence of operations is to write the binary expansion of $m-1$. Then each 0 in this expansion corresponds to a $\ominus$ operation and each 1 corresponds to a $\oplus$ operation. Using this binary labeling, we define a binary stochastic process. Assume that $\Omega = \{0,1\}^{\infty}$, and $\mathcal{F}$ is the $\sigma$-algebra generated by the cylindrical sets

$$S_{(i_1,i_2,\ldots,i_s)} = \{\omega \in \Omega \text{ such that } \omega_{i_1} = 1, \ldots, \omega_{i_s} = 1\}$$

for every integer $s$ and $i_1, i_2, \ldots, i_s$. We also define $\mathcal{F}_n$ as the $\sigma$-algebra generated by the first $n$ coordinates of $\omega$. In other words, $\mathcal{F}_n$ is the $\sigma$-algebra generated by sets of the form

$$\{\omega \in \Omega \text{ such that } \omega_1 = 1, \ldots, \omega_n = 1\}.$$

Let $\mathcal{F}_0 = \{\emptyset, \Omega\}$ be the trivial $\sigma$-algebra. We also define the uniform probability measure $\mu$ over the cylindrical sets by

$$\mu(S_{(i_1,i_2,\ldots,i_n)}) = \frac{1}{2^n},$$

which by uniformity assumption, is independent of the values taken by $i_1, i_2, \ldots, i_n$. This measure can be uniquely extended to $\mathcal{F}$. Let $[\omega]_n = \omega_1\omega_2\ldots\omega_n$ denote the first $n$ coordinates of $\omega = \omega_1\omega_2\ldots$ and $Y_{[\omega]_n}$ denote the random variable $Y_i$, where the binary expansion of $i-1$ is $[\omega]_n$, and let $Y^{[\omega]_n}$ denote

$$Y^{[\omega]_n} = \{Y_{[\eta]_n} : \ \eta < \omega\}.$$

We also define the random variable $I_n$ by

$$I_n(\omega) = H(Y_{[\omega]_n} | Y^{[\omega]_n}). \tag{5}$$

As an example, if $\omega = 0.10\ldots$ then

$$I_2(\omega) = H(Y_{10}|Y_{01}, Y_{00}) = H(Y_3|Y_1, Y_2).$$

It is also important to note that

$$I_{n+1}([\omega]_n, 0) = H(Y_{[\omega]_n} + \tilde{Y}_{[\omega]_n} | Y^{[\omega]_n}, \tilde{Y}^{[\omega]_n}) \tag{6}$$

where $\tilde{}$ denotes an independent copy of the corresponding random element.

**Theorem 3.** *$(I_n, \mathcal{F}_n)$ is a martingale.*

*Proof:* $I_n$ is adapted to $\mathcal{F}_n$ by definition. Hence it is sufficient to show that $E\{I_{n+1}|\mathcal{F}_n\} = I_n$. For simplicity, we prove the case $n = 1$. The general case is similar. Using Figure 2, we have

$$\begin{aligned} T(\omega_1) &= E\{I_2(\omega)|\mathcal{F}_1\} \\ &= E\{I_{\omega_1\omega_2}|\mathcal{F}_1\} \\ &= \frac{1}{2}(I_{\omega_1 0} + I_{\omega_1 1}), \end{aligned}$$

which is a function of $\omega_1$.

$$\begin{aligned} T(0) &= \frac{1}{2}(I_{00} + I_{01}) \\ &= \frac{1}{2}(H(Y_{00}) + H(Y_{01}|Y_{00})) \\ &= \frac{1}{2}H(Y_{00}, Y_{01}) = \frac{1}{2}H(Y_0, \tilde{Y}_0) = H(Y_0). \end{aligned}$$

We can also show that $T(1) = H(Y_1|Y_0)$. Hence, $T(\omega_1) = I_1(\omega)$ and $E\{I_2|\mathcal{F}_1\} = I_1$. Similarly, we can show that $E\{I_{n+1}|\mathcal{F}_n\} = I_n$. ■

## C. Main Theorem

In this section, we prove the main theorem of the paper.

*Proof of Theorem 1:* Assume that $Y_1^N = G_N X_1^N$, for $N = 2^n, n \in \mathbb{Z}_+$, and $H_i = H(Y_i|Y_1^{i-1})$, $i \in [N]$. Also fix $\epsilon > 0$. Let us define

$$\begin{aligned} K_n &= \{i : i \in [N], H_i > \epsilon\}, \\ Y_{[K_n]} &= \{Y_j : \ j \in [K_n]\}. \end{aligned}$$

Hence, by Definition 3, $|K_n| = m_N$ and $\bar{J}_N$ is obtained from $J_N$ by selecting the rows with index $K_n$. We have

$$\begin{aligned} H(X_1^N|\bar{J}_N X_1^N) &= H(X_1^N) - I(X_1^N; \bar{J}_N X_1^N) \\ &= H(Y_1^N) - H(Y_{[K_n]}) \\ &= H(Y_{[K_n^c]}|Y_{[K_n]}) \\ &\le \sum_{i \in K_n^c} H(Y_i|Y_1^{i-1}) \\ &\le |K_n^c|\epsilon = (N - m_N)\epsilon, \end{aligned}$$

which implies that

$$\frac{H(X_1^N|\bar{J}_N X_1^N)}{N} \le \frac{(N - m_N)\epsilon}{N} \le \epsilon.$$

This shows that the family $\{\bar{J}_N\}$ is $\epsilon$-REP. Now it remains to show that the measurement rate of this family is 0. To prove this, we construct the martingale $I_n$ by (5). $I_n$ is a positive martingale and converges to a random variable $I_\infty$ almost surely. Our aim is to show that for any two positive numbers $a$ and $b$ where $a < b$, $\mu(I_\infty \in (a,b)) = 0$. which implies that $\mu(I_\infty \in \{0, \infty\}) = 1$. Since $I_n$ is a martingale, $E\{I_n\} = E\{I_0\} = H(X) < \infty$. Using Fatou's lemma we obtain

$$E\{I_\infty\} \le \liminf E\{I_n\} = H(X_1) < \infty,$$

which implies that $\mu(I_\infty = \infty) = 0$. Hence, $I_n$ converges almost surely to 0 and it also converges to 0 in probability. In other words, given $\epsilon > 0$,

$$\limsup_{n\to\infty} \mu(I_n > \epsilon) = \limsup_{n\to\infty} \frac{|K_n|}{2^n} = \limsup_{N\to\infty} \frac{m_N}{N} = 0.$$

This implies that for a fixed $\epsilon > 0$ the measurement rate $\rho$ is 0. Now it remains to prove that for any two positive numbers $a$ and $b$, where $a < b$, $\mu(I_\infty \in (a, b)) = 0$. Fix a $\delta > 0$ then for every $\omega$ in the convergence set there is a $n_0$ such that for $n > n_0$, $|I_{n+1}(\omega) - I_n(\omega)| < \delta$. Using the martingale proprty

$$I_n(\omega) = \frac{1}{2}(I_{n+1}([\omega]_n, 0) + I_{n+1}([\omega]_n, 1)),$$

we obtain that for $n > n_0$,

$$|I_{n+1}(\omega) - I_n(\omega)| = |I_{n+1}([\omega]_n, 0) - I_n([\omega]_n)| < \delta.$$

Using (6) and the entropy power inequality (3), it results that $0 \le I_n(\omega) < \rho(\delta)$ where $\rho(\delta)$ can be obtained from $g$. This implies that $I_n$ must converge to 0. ■

## VI. Numerical Simulations

For simulation, we use a binary random variable, where $p_X(0) = 1 - p$ for some $0 < p \le \frac{1}{2}$.

### A. Absorption Phenomenon

Figure 3 shows the absorption phenomenon for $p = 0.05$ and $N = 64, 128, 256, 512$.

### B. Nested Property

Absorption phenomenon is shown in Figure 4 for $N = 512$ and different values of $p$. It is seen that the high entropy indices for smaller $p$ are *included* in the high entropy indices of the larger one. We call this the *"nested"* property. The benefit of the nested property is that it allows one to take measurements adaptively if the sparsity level is unknown. In other words, one takes some measurements corresponding to the high entropy indices and if the recovery is not successful, refines them by adding extra measurements that correspond to the indices with lower entropy to improve the quality of recovery.

### C. Robustness to Measurement Noise

Figure 5 shows the stability analysis of the reconstruction algorithm to Gaussian measurement noise. For simulation, we used $N = 512$, $p = 0.05$ and took all of the indices with entropy greater than 0.01. In other words, the measurement matrix was 0.01-REP for the binary distribution $p$. For recovery, we use ML decoder which exploits the recursive structure of the polar code. Let denote the input random variables by $X_1^N$ and assume that we keep all of the rows of the matrix $J_N$ with indices in the set $K$. We define the SNR( signal to noise ratio) at the input of the decoder as:

$$\text{SNR}_{\text{in}} = \frac{|K|\,\sigma^2}{\sum_{i\in K} E(Y_i^2)},$$

and the SNR at the output of the decoder as

$$\text{SNR}_{\text{out}} = \frac{1}{N}\sum_{i=1}^{N} E(|X_i - \hat{X}_i|^2),$$

where $\sigma^2$ is the noise variance and $\hat{X}_i$ is the output of the ML decoder. The result shows approximately 4 dB loss in SNR for high SNR regime. Notice that some part of this loss results from the finite distortion 0.01 that we tolerate by removing the measurements corresponding to low entropy indices.

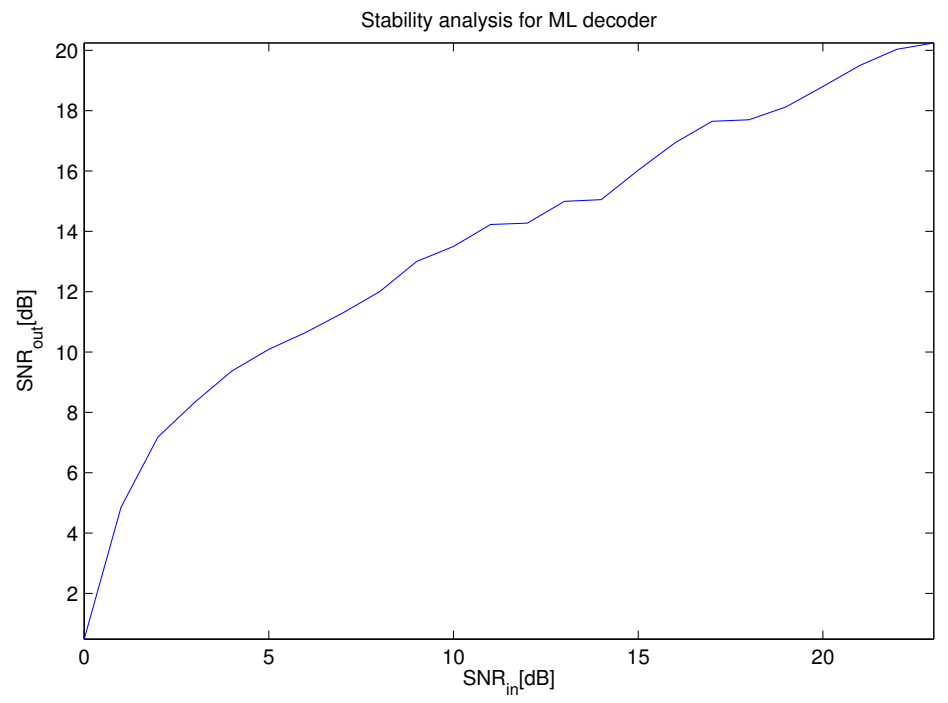

Fig. 5: Stability analysis

## Appendix

## Proof of Lemma 1

Let $X_1^N$ be a set of $N$ i.i.d. random variables with a uniform distribution over $[-1, 1]$. Let $D_i = Q(X_i)$, $i \in [N]$, be the uniform quantizer output for $X_i$. It is easy to see that we can write $X_i = \frac{2D_i - q + 1}{q} + C_i$, $i \in [N]$, where $C_i$ is the quantization noise which is uniformly distributed over $[-\frac{1}{q}, \frac{1}{q}]$. Moreover, $C_i$ is independent of $D_i$. As $\Phi_N$ is full rank, the vector random variable $\Phi_N X_1^N$ has a well-defined density over $\mathbb{R}^{m_N}$. We have

$$\begin{aligned} &H(D_1^N|\Phi_N X_1^N) \\ &= H(D_1^N) - I(D_1^N; \Phi_N X_1^N) \\ &= N\log_2(q) - h(\Phi_N X_1^N) + h(\Phi_N X_1^N|D_1^N) \\ &= N\log_2(q) - h(\Phi_N X_1^N) + h(\Phi_N C_1^N) \\ &= N\log_2(q) - h(\Phi_N X_1^N) + h(\Phi_N X_1^N) - m_N\log_2(q) \\ &= (N - m_N)\log_2(q) < N\epsilon, \end{aligned}$$

where $I$ denotes the mutual information between two random variables and $h$ is the differential entropy for continuous distributions. This implies that

$$\rho = \limsup_{N\to\infty} \frac{m_N}{N} \ge 1 - \frac{\epsilon}{\log_2(q)},$$

which gives the desired result as $\epsilon$ goes to 0. In the proof we used the fact that $\Phi_N X_1^N$ has the same distribution as $q \times \Phi_N C_1^N$.

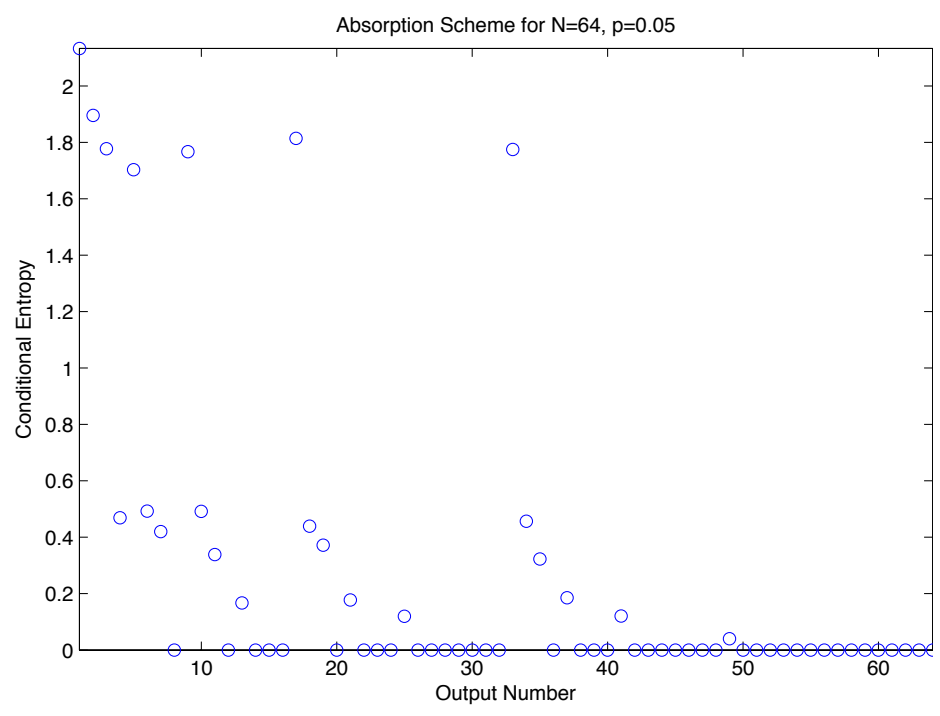

(a) $N = 64$

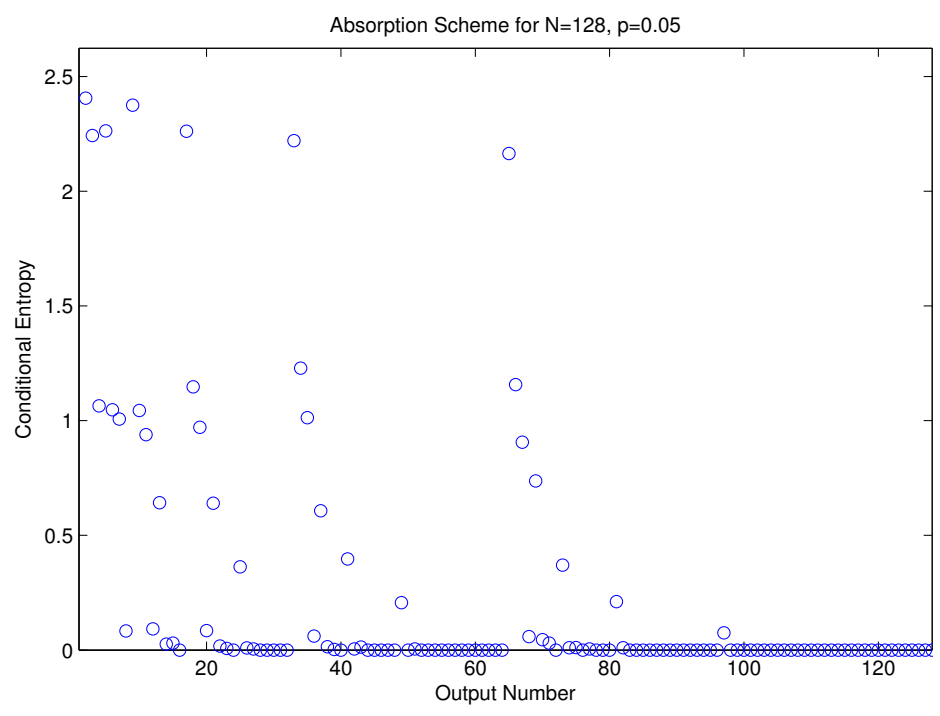

(b) $N = 128$

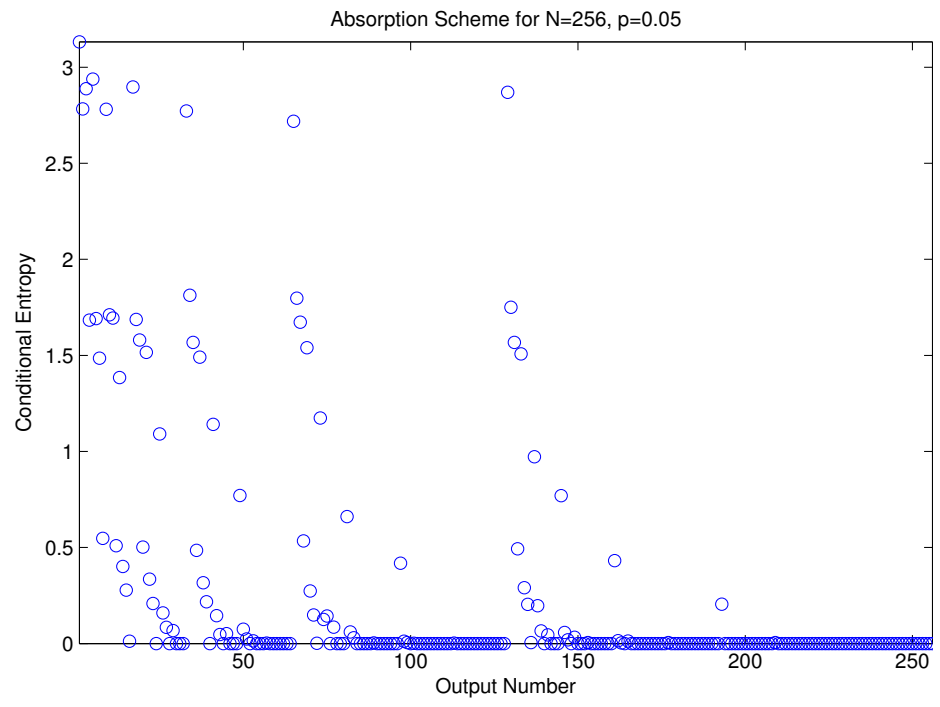

(c) $N = 256$

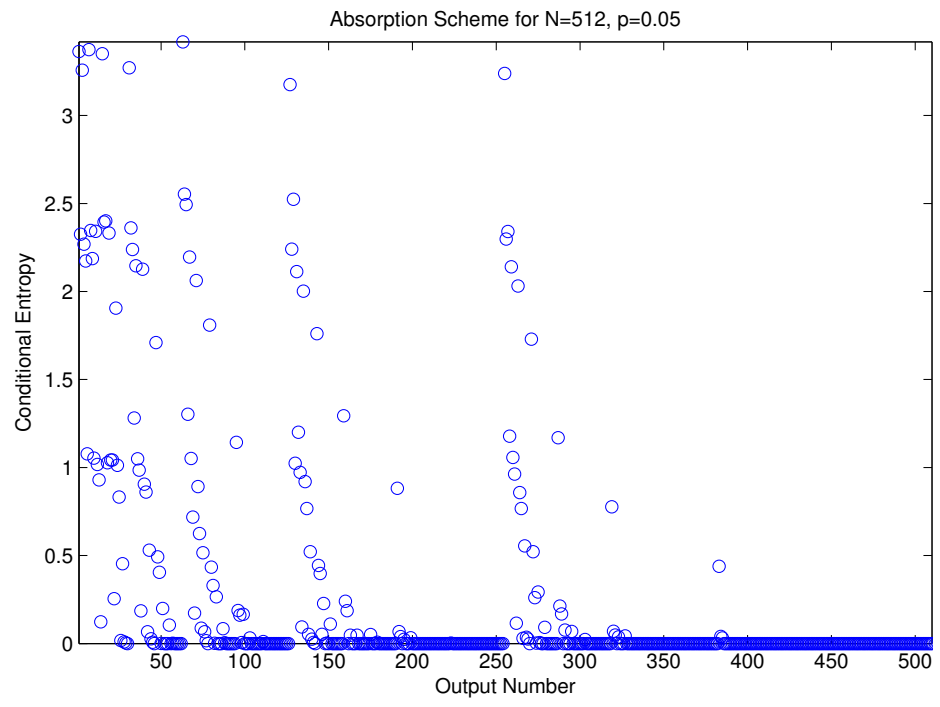

(d) $N = 512$

Fig. 3: Absorption trace for $p = 0.05$

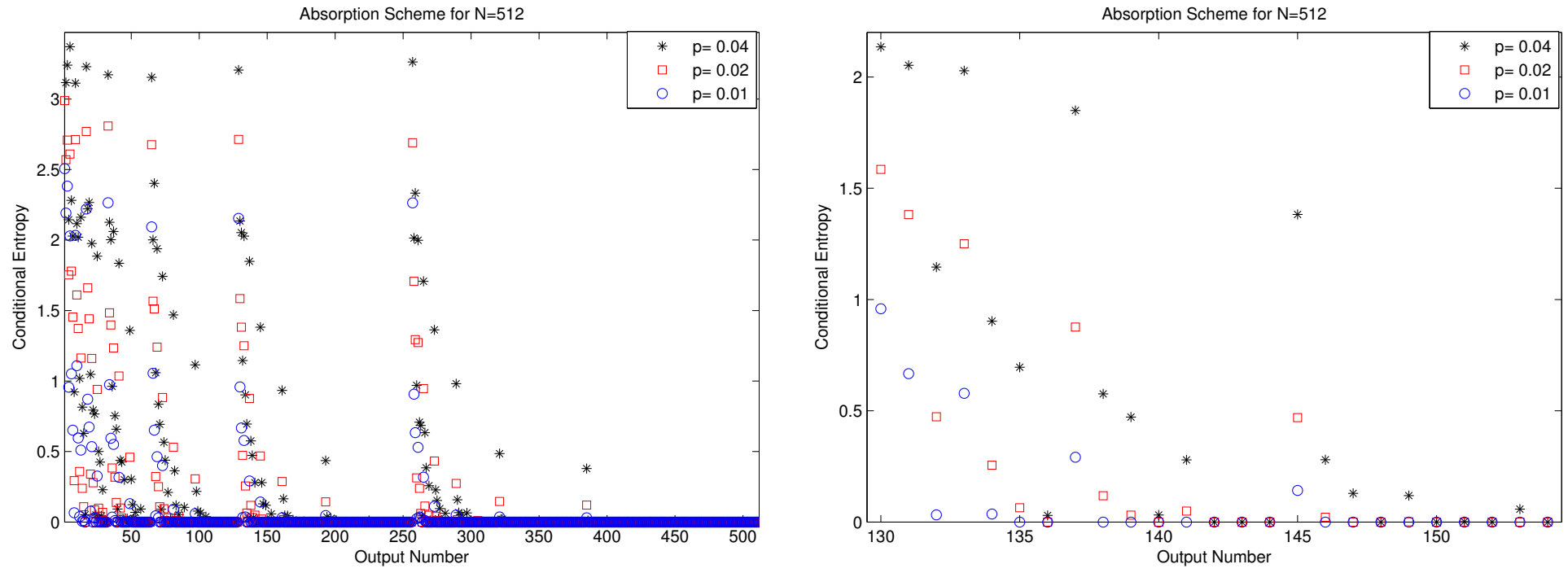

Fig. 4: Nested property for $N = 512$ and different $p$